\documentclass[12pt]{article}
\usepackage{amsfonts}
\usepackage{latexsym,cite,amssymb}
\usepackage{times}
\usepackage{graphicx}
\usepackage{color}
\makeatletter
\renewcommand\section{\@startsection {section}{1}{\z@}%
                                   {-3.5ex \@plus -1ex \@minus -.2ex}
                                   {2.3ex \@plus.2ex}%
                                   {\normalfont\large\bfseries}}

\renewcommand\subsection{\@startsection{subsection}{2}{\z@}%
                                     {-3.25ex\@plus -1ex \@minus -.2ex}%
                                     {1.5ex \@plus .2ex}%
                                     {\normalfont\bfseries}}

\renewcommand{\theequation}{\thesection.\arabic{equation}}

\makeatother

\newcommand{\bea}{\begin{eqnarray}}
\newcommand{\eea}{\end{eqnarray}}
\newcommand{\be}{\begin{equation}}
\newcommand{\ee}{\end{equation}}
\newcommand{\bem}{\begin{pmatrix}}
\newcommand{\eem}{\end{pmatrix}}
\newcommand{\bl}{\begin{align}}
\newcommand{\el}{\end{align}}

\begin{document}
\begin{center}
${}$ \thispagestyle{empty}

\vskip 3cm {\LARGE {\bf Analytical study on holographic
superconductors in external magnetic field}} \vskip 1.25 cm  {
Xian-Hui Ge${}^{a}~$\footnote{ gexh@shu.edu.cn},
  Bin Wang${}^{b}~$\footnote{
wangb@fudan.edu.cn}, Shao-Feng Wu${}^{a}~$\footnote{
sfwu@shu.edu.cn}, Guo-Hong Yang${}^{a}~$\footnote{
ghyang@shu.edu.cn} }
 ~~~
\vskip 0.5cm${}^{a}$Department of Physics, Shanghai University, Shanghai 200444, China\\
~{}
${}^{b}$Department of Physics, Shanghai Jiao Tong University, Shanghai 200240, China\\
~{}
\\

~~~\\
~~~\\

\vspace{1cm}

\begin{abstract}
\baselineskip=16pt

We investigate the holographic superconductors immersed in an
external magnetic field by using the analytical approach.  We obtain
the spatially dependent condensate solutions in the presence of the
magnetism and find analytically that the upper critical magnetic
field satisfies the relation given in the Ginzburg-Landau theory.
The external magnetic field expels the condensate and makes the
condensation harder to form. Extending to the D-dimensional
Gauss-Bonnet AdS black holes, we examine the influence given by the
Gauss-Bonnet coupling on the condensation. Different from the
positive coupling, we find that the negative Gauss-Bonnet coupling
enhances the condensation when the external magnetism is not strong
enough.
\end{abstract}
\end{center}

\newpage

\section{Introduction }
The AdS/CFT correspondence \cite{ads/cft,gkp,w}, which has been
proved as one of the most fruitful ideas in string theory, describes
that a string theory on asymptotically AdS spacetimes can be related
to a conformal field theory on the boundary. Using this
gauge/gravity correspondence, Gubser first suggested that the
spontaneous $U(1)$ symmetry breaking by bulk black holes can be used
to construct gravitational duals of the transition from normal state
to superconducting state in the boundary theory \cite{gub1,gub2}.
This investigation is not easy, since the full equations are coupled
and nonlinear, and so sophisticated numerical methods and some
limits have to be employed in order to extract the key physics.
Gubser studied the case of non-abelian Reissner-Nordstrom black
holes with gauge field $A=A_tdt$ which influences the effective mass
of the scalar and contributes to its condensing. In a simpler model,
Hartnoll et al. considered a neutral black hole with a charged
scalar and the only Maxwell sector $A=A_t$ and they captured the
essence in this limit and showed that the properties of a
(2+1)-dimensional superconductor can indeed be reproduced
\cite{horowitz}. This study has been further extended by
investigating how the condensate behaves when the external magnetic
field is added to the system \cite{horowitz2,aj,amm,john,mon}, where
the spatially dependent Maxwell sectors have to be considered in the
full solutions. Counting on the numerical calculations,  the droplet
solution \cite{aj,john} and the vortex configuration
\cite{john,mon,ns} have been found for the holographic
superconductor in the presence of magnetic field. Along this line,
the application of the AdS/CFT correspondence to condensed matter
physics has been widely studied
\cite{Hartnoll,h08,wen,ama,kou,maeda,son,zeng,Sin,cz,cj,sin,wcj,siop}(see
\cite{hart,herzog} for reviews) and recently a further progress
beyond the probe limit by considering the backreaction of the scalar
field on the background spacetime has been reported in
\cite{horowitz3}.

Motivated by the application of the Mermin-Wagner theorem to the
holographic superconductors, recently there were studies of the
effects of the curvature corrections on the (3+1)-dimensional
superconductor \cite{gre,pw}. By considering the model of a charged
scalar field together with a Maxwell field with $A=A_tdt$ in the
high-dimensional Gauss-Bonnet-AdS black hole backgrounds, it was
found that the bigger positive Gauss-Bonnet coupling which reflects
higher curvature correction makes condensation harder. A
semi-analytical method was introduced in understanding the
condensation \cite{gre} and was further refined in \cite{pw}. This
semi-analytical method can explain the qualitative features of
superconductors and gives fairly good agreement with numerical
results.

In this work, we will apply the analytical method developed in
\cite{gre,pw} to investigate the holographic superconductor in the
presence of the external magnetic field in the probe approximation.
We hope that the analytic investigation can help us pick out more
physics in a straightforward way. From the Ginzburg-Landau theory,
we know that the upper critical magnetic field has the well-known
form \cite{poole} \be
B_{c2}=\frac{\Phi_0}{2\pi\xi(T)^2}=\frac{\Phi_0}{2\pi
\xi(0)^2}\left(1-T/T_c\right), \ee where $\Phi_0$ and $\xi(T)$
denote the quantum flux and the Ginzburg-landau coherent length,
respectively. By employing the semi-analytic method,  we will show
that we can reproduce the relation $B_{c2}\propto (1-T/T_c)$
analytically for the holographic superconductor.

We will further extend our investigation to the Gauss-Bonnet higher
dimensional black holes and generalize the previous study
\cite{gre,pw} by adding magnetic charge to the black hole and
immersing the superconductor into an external magnetic field. The
Gauss-Bonnet coupling constant is constrained simultaneously by the
positivity of the energy constraints in conformal field theories
\cite{hofman} and causality in their dual gravity
description\cite{brig}. There is an upper positive bound for the
Gauss-Bonnet coupling and beyond which the tensor type perturbation
at the boundary would propagate at a superluminal velocity and the
boundary theory would then become
pathologic\cite{gsst,ge,ges,bm,cai,shu,boe,esc,ce}. For the positive
Gauss-Bonnet coupling below this upper bound, it was observed in
general that the bigger coupling makes the condensation harder to
form\cite{gre,pw}. Besides there also exists a lower bound on the
Gauss-Bonnet coupling by considering the causality\cite{bm,ce}. In
general for dimension $D\geq 5$, the Gauss-Bonnet coupling is
bounded within the range \be \label{range}
-\frac{(D-3)(3D-1)}{4(D+1)^2}\leq \lambda\leq
\frac{(D-3)(D-4)(D^2-3D+8)}{4(D^2-5D+10)^2}. \ee In addition to
checking the qualitative property observed in \cite{gre,pw} for the
influence of the positive Gauss-Bonnet coupling on the
superconductor in the presence of the external magnetic field, we
will further examine the allowed negative coupling influence on the
condensation.

We have three coupled nonlinear partial differential equations
involving the scalar field $\psi$, the scalar potential $A_t$ and
vector potential $\textbf{A}$, which are even more complicated than
the Ginzburg-Landau equations. To solve these equations analytically
we will follow the logic used by Abrikosov \cite{abri} as listed in
table I. At first we will start to consider the weak magnetic field
limit so that $\textbf{A}\sim 0$ and obtain the spatially
independent condensate solutions by using the method given in
\cite{gre,pw}.  To the sub-leading order, we can treat the magnetic
field as a small perturbation. Our main purpose in this work is to
calculate the upper critical magnetic field $B_{c2}$ associated with
the holographic superconductors in the backgrounds of
$4$-dimensional Schwarzschild-AdS and $D$-dimensional
Schwarzschild-AdS-Gauss-Bonnet black holes. So secondly we will
consider that the magnetic field is strong enough. We will regard
the scalar field $\psi$ as a perturbation and examine its behavior
in the presence of strong magnetism.  In this case, we will seek
non-trivial spatially dependent solutions of condensation.

\begin{table*}[htbp]\label{table}
\begin{center}
\begin{tabular}{|c|c|c|c|c|c|c|c|c|}
\hline
$$&$\rm weak~~~ magnetic~~~ field$&$\rm strong~~~ magnetic~~~ field$\\
\hline
$\psi$&$  \psi=\psi(z) $&$  \psi=\psi(z,x,y), \psi\sim 0$\\
\hline
$A_t$&$  A_t=A_t(z) $&$ A_t=A_t(z) $\\
\hline
$A_{\varphi}$&$  A_{\varphi}= A_{\varphi}(z,x,y),A_{\varphi}\sim 0$&$  A_{\varphi}= A_{\varphi}(z) $\\

\hline
\end{tabular}
\caption{Logic of the analytic calculation. We will work in two
limits respectively: weak magnetic field limit and strong magnetic
field limit.}\label{table}
\end{center}
\end{table*}
The paper is organized as follows: In section 2, we will study the
$(2+1)$-dimensional holographic superconductors immersed in the
external magnetic field by using the semi-analytic methods proposed
in \cite{gre,pw} and obtain the expression for the upper critical
magnetic field. In section 3, we will extend our investigation to
Gauss-Bonnet black hole backgrounds in $D$ dimensions. The
conclusions and discussions will be provided in the last section.

\section{$(2+1)$-dimensional holographic superconductors immersed in an external magnetic field}

\subsection{The background}

We begin with the 4-dimensional Schwarzschild AdS black hole with
the metric
\begin{equation}
ds^2=\frac{r^2}{l^2}\left(-f(r)dt^2+\sum^{2}_{i}dx_{i}^2\right)+\frac{l^2}{r^2f(r)}dr^2,
\end{equation}
where the metric coefficient
\begin{equation}
f(r)=1-\frac{Ml^2}{r^3}=1-\frac{r^3_{+}}{r^3},
\end{equation}
and $l$ is the AdS radius and $M$ is the mass of the black hole. The
Hawking temperature of the black hole is $T=\frac{3M^{1/3}}{4\pi
L^{4/3}}$. Setting $z=\frac{r_{+}}{r}$, the metric can be rewritten
in the form
\begin{equation}
ds^2=\frac{l^2
\alpha^2}{z^2}\left[-f(z)dt^2+dx^2+dy^2\right]+\frac{l^2}{z^2
f(z)}dz^2,
\end{equation}
where
\begin{equation}
f(z)=1-z^3,~~~\alpha=\frac{r_{+}}{l^2}=\frac{4}{3}\pi T.
\end{equation}
We introduce a charged, complex scalar field into the 4-dimensional
Einstein-Maxwell action with negative cosmological constant
\begin{equation}\label{action}
S=\frac{1}{16\pi G_{4}}\int d^4
x\sqrt{-g}\bigg\{R-2\Lambda-\frac{1}{4}F_{\mu\nu}F^{\mu\nu}
-|\partial_{\mu}\psi-iA_{\mu}\psi|^2-m^2|\psi|^2\bigg\},
\end{equation}
where $G_{4}$ is the 4-dimensional Newton constant, the cosmological
constant $\Lambda=-3/l^2$ and
$F_{\mu\nu}=\partial_{\mu}A_{\nu}-\partial_{\nu}A_{\mu}$. In the
probe approximation, the Maxwell and scalar field equations obey
\begin{eqnarray}
&&z^2 \partial_{z}\left(\frac{f(z)}{z^2}\partial_{z}
\psi\right)+\frac{A^2_t}{\alpha^2
f(z)}\psi-\frac{m^2l^2}{z^2}\psi\nonumber\\
&&=\frac{-1}{\alpha^2}\left[(\partial_{x}-iA_x)^2+
(\partial_{y}-iA_y)^2\right]\psi,\label{main1}\\
&&
\left(\alpha^2f(z)\partial^2_{z}+\partial^2_{x}+\partial^2_{y}\right)A_t=\frac{2l^2\alpha^2}{z^2}A_t|\psi|^2,\label{at}\\
&& \left(\partial_{z}\alpha^2
f(z)\partial_{z}+\partial^2_{x}+\partial^2_{y}\right)A_i-\partial_i(\delta^{jk}\partial_j
A_k)=-\frac{l^2 \alpha^2}{z^2}j_i,
\end{eqnarray}
where $i,j,k=1,2,3$ and $j_i=i(\psi
\partial_{i}\psi^{*}-\psi^{*}\partial_{i}\psi)+2A_i |\psi|^2$.
Considering the properties of AdS spacetimes, we can impose the
following boundary conditions:\\ 1) At the asymptotic AdS boundary
($z\rightarrow 0$), the solution of the scalar field behaves like
\begin{equation}
\psi\sim c_1z^{\triangle_{-}}+c_2z^{\triangle_{+}},
\end{equation}
where $\triangle_{\pm}=\frac{3}{2}\pm \sqrt{\frac{9}{4}+m^2l^2}$ and
the coefficients $c_1$ and $c_2$ both multiply normalizable modes of
the scalar field equations and according to the AdS/CFT
correspondence, they correspond to the vacuum expectation values
$c_1=<\mathcal {O}_{-}>$ and $c_2=<\mathcal {O}_{+}>$ of an operator
$\mathcal {O}$ dual to the scalar field. Here we consider the case
by setting $c_1=0$ and $m^2l^2=-2$ (i. e. $\triangle_{+}=2$ )which
corresponds to the faster falloff dual to the expectation
value for simplicity. \\
 2) The asymptotic values of the Maxwell field at the AdS boundary give the chemical
 potential and the external magnetic field
\begin{equation}
\mu=A_{t}(\textbf{x}, z\rightarrow
0),~~~B(\textbf{x})=F_{xy}(\textbf{x},z\rightarrow 0).
\end{equation}
3) The boundary condition at the horizon requires  $\psi$ and
$A_i(\textbf{x},z=1)$ regular and $A_{t}(\textbf{x},z=1)=0$.

\subsection{Weak magnetic field limit}
In the weak magnetic field limit, it is consistent to consider the
scalar field $\phi$ and $A_t$ as functions of $z$ only to the
leading order. Thus it is easy to derive the spatially independent
condensate solution that corresponds to the superconductor phase
below some critical temperature by employing the analytic method
developed in \cite{gre,pw}. This analytic solution will be helpful
in our understanding on the magnetic induced currents in the
superconducting phase.

\subsubsection{Condensation}
In the weak magnetic field limit, $A_i \sim 0$, the equations of
motion reduce to
\begin{eqnarray}
&&z^2 \partial_{z}\left(\frac{f(z)}{z^2}\partial_{z}
\psi\right)+\frac{A^2_t}{\alpha^2
f(z)}\psi-\frac{m^2l^2}{z^2}\psi=0,\label{main11}\\
&&
\alpha^2f(z)\partial^2_{z}A_t=\frac{2l^2\alpha^2}{z^2}A_t|\psi|^2.\label{mat}
\end{eqnarray}
The boundary condition at the horizon $z=1$ reads,
\begin{equation}\label{215}
A_t=\phi(z)=0, ~~~\psi'(1)=\frac{2}{3}\psi(1),
\end{equation}
and near the asymptotic AdS boundary $z\rightarrow 0$ we have
\begin{equation}
\phi(z)=\mu-\frac{\rho}{r_{+}}z,~~~\psi=c_2z^2.
\end{equation}
Matching the asymptotic solutions near the horizon and the AdS
boundary at the intermediate point, say $z_{m}=1/2$, we get
\begin{eqnarray}\label{217}
&&\psi(1)=\frac{\sqrt{3}}{l}\sqrt{\frac{\rho}{-\phi^{'}(1)r_{+}}}
\sqrt{1-\frac{-\phi^{'}(1)r_{+}}{\rho}},~~~c_2=\frac{5}{3}\psi(1),\label{3.5}
\\&&-\phi^{'}(1)=2\sqrt{7}\alpha.
\end{eqnarray}
Finally the expectation value of the 2-dimensional operator
$<\mathcal {O}_2>=\sqrt{2}c_2 r^2_{+}/l^3$ can be obtained in the
form
\begin{equation}\label{o2}
<\mathcal
{O}_2>=\frac{80\pi^2}{9}\sqrt{\frac{2}{3}}T_cT\sqrt{1+\frac{T}{T_c}}\sqrt{1-\frac{T}{T_c}},
\end{equation}
where the critical temperature is defined by
\begin{equation}\label{critical}
T_c\left(B\sim 0\right)\equiv
\frac{T\sqrt{\rho}}{l\sqrt{-\phi^{'}(1)\alpha}}=\frac{3\sqrt{\rho}}{4\pi
l\sqrt{2\sqrt{7}}}.
\end{equation}
In the presence of a strong magnetic field, we will show that the
critical temperature is dependent on the magnetic field $B$ and
there is an upper bound on the magnetic field for the condensation.
We will see in the next section that when we gradually cool a
superconductor in an external field, at the zero field transition
temperature, $T_c(B\sim 0)$, it will be impossible to condensate
with nonzero $B$.

Before going further, we would like to comment (\ref{o2}). Although
it was found that the analytic approach can explain the qualitative
features of superconductors and agrees fairly well with numerical
results when $T\sim T_c$, it breaks down when $T\rightarrow 0$. This
breakdown of the analytic method will also been seen in the presence
of strong magnetism in section 3.

\subsection{Strong magnetic field limit}
In this subsection, we will explore the effect of a strong external
magnetic filed on the holographic superconductors and seek
non-trivial spatially dependent condensate solutions.  We will
regard the scalar field $\psi$ as a perturbation and examine its
behavior in the neighborhood of the upper critical magnetic field
$B_{c2}$. In this case, $\psi$ is a function of the bulk coordinate
$z$ and the boundary coordinates $(x,y)$ simultaneously. According
to the AdS/CFT correspondence, if the scalar field $\psi\sim
X(x,y)R(z)$, the vacuum expectation values $<\mathcal {O}> \propto
X(x,y)R(z)$ at the asymptotic AdS boundary (i.e. $z\rightarrow
0$)\cite{aj,ns}. One can simply write $<\mathcal {O}> \propto R(z)$
by dropping the overall factor $X(x,y)$. To the leading order, it is
consistent to set the ansatz
\begin{equation}
\psi=\psi(x,y,z),~~~A_t=\phi(z),~~~A_x=0,~~~A_y=B_{c2}x.
\end{equation}
The equation of motion for $\psi$ then becomes
\begin{equation}
z^2 \partial_{z}\left(\frac{f(z)}{z^2}\partial_{z}
\psi\right)+\frac{\phi^2}{\alpha^2 f(z)}\psi-\frac{m^2l^2}{z^2}\psi
 =\frac{-1}{\alpha^2}\left[\partial_{x}^2+
(\partial_{y}-iB_{c2}x)^2\right]\psi.
\end{equation}
This equation can be solved by separating the variables
\begin{equation}
\psi=e^{ik_y y}X(x)R(z),
\end{equation}
where $X(x)$ is governed by the equation of a two dimensional
harmonic oscillator with frequency determined by $B_{c2}$,
\begin{equation}
\label{hermite1}
\frac{1}{\alpha^2}\left[-X''(x)+(k_y-B_{c2}x)^2X\right]=\frac{\lambda_n
B_{c2}}{\alpha^2}X(x),
\end{equation}
and $R(z)$ satisfies
\begin{equation}\label{Rz}
f(z)R''(z)-\left(\frac{2f(z)+3z^3}{z}\right)R'(z)+\frac{
\phi^2}{\alpha^2 f(z)}R(z)+\frac{2}{z^2}R(z)
 =\frac{\lambda_n
B_{c2}}{\alpha^2}R(z),
\end{equation}
where $\lambda_n=2n+1$ denotes the separation constant. The solution
of (\ref{hermite1}) is nothing but the Hermite functions $H_n$
\begin{equation}
X(x)=e^{-{\frac{(B_{c2}x-k_y)^2}{2 B_{c2}}}} H_n(x).
\end{equation}
One may pay attention to the lowest mode $n=0$, which is the first
to condensate and is the most stable solution after condensation.

(\ref{Rz}) can be solved by using the analytical method proposed in
\cite{gre,pw}. Keeping in mind that regularity at the horizon $z=1$
requires
\begin{equation}
\label{R1} R'(1)=\frac{2}{3}R(1)-\frac{ B_{c2}}{3\alpha^2}R(1),
\end{equation}
and near the AdS boundary $z\rightarrow 0$ it sets
\begin{equation}\label{c2}
R(z)=c_1 z+c_2 z^2.
\end{equation}
We will choose $c_1=0$ for simplicity. The scalar potential
satisfies the boundary condition at the asymptotic AdS region
$A_t=\phi(z)=\mu-\frac{\rho}{r_+}z$ and vanishes at the horizon
$A_t=0,$ as $z\rightarrow 1$. In the strong field limit, the scalar
field $\psi$ is almost vanishing. In this sense, we can drop out the
$|\psi|^2$ term in the right hand side of equation (\ref{at}). It is
easy to find that $A_t=\phi(z)=\frac{\rho}{r_+} (1-z)$ is a solution
that satisfies (\ref{at}) and the corresponding boundary conditions
\cite{ns}. In the following calculation, we will use the ansatz for
$A_t$ to compute the relation between the critical temperature
$T_c(0)$ and the upper critical magnetic field $B_{c2}$.

Expanding $R(z)$ in a Taylor series near the horizon, we obtain,
\begin{equation}\label{exp}
R(z)=R(1)-R'(1)(1-z)+\frac{1}{2}R''(1)(1-z)^2+\ldots.
\end{equation}
Near $z=1$, (\ref{Rz}) gives
\begin{equation}
\label{R2}
R''(1)=-\frac{2}{3}R'(1)-\frac{\phi~'(1)^2}{18\alpha^2}R(1)-\frac{
B_{c2}}{6\alpha^2}R'(1)-\frac{ B_{c2}}{3\alpha^2}R(1).
\end{equation}
Substituting (\ref{R1}) and (\ref{R2}) into (2.64), we find the
approximate solution near the horizon
\begin{equation}\label{exp}
R(z)=\frac{1}{3}R(1)+\frac{2}{3}R(1)z+\frac{
B_{c2}}{3\alpha^2}R(1)(1-z)-\left(\frac{2}{9}+\frac{\phi~'(1)^2}{36
\alpha^2}+\frac{ B_{c2}}{9 \alpha^2}-\frac{
B^2_{c2}}{36\alpha^4}\right)R(1)(1-z)^2,
\end{equation}where $R^2$ terms have been neglected.
Now let us match the solutions (\ref{exp}) and (\ref{c2}) at the
intermediate point $z_m=1/2$. Requiring the solutions to be
connected smoothly, we obtain
\begin{eqnarray}
&&36 \alpha^4 {c_2}=\left(88 \alpha^4+{20 \alpha^2 B_{c2}}-\alpha^2
{\phi~'(1)^2}+{
B^2_{c2}}\right)R(1), \\
&&36 \alpha^4 {c_2}=\left(32 \alpha^4+\alpha^2
{\phi~'(1)^2}-8\alpha^2{ B_{c2}}-{ B^2_{c2}}\right)R(1),
\end{eqnarray}
which leads
\begin{equation}\label{BT}
 {\phi~'(1)^2}=28\alpha^2+14{ B_{c2}}+\frac{
B^2_{c2}}{\alpha^2}. \ee When $B_{c2}=0$, $\phi'(1)$ goes back to
(2.18), which leads to the critical temperature for the condensation
exhibited in (\ref{critical}).
\begin{figure}[htbp]
\begin{center}

\includegraphics*[scale=0.7] {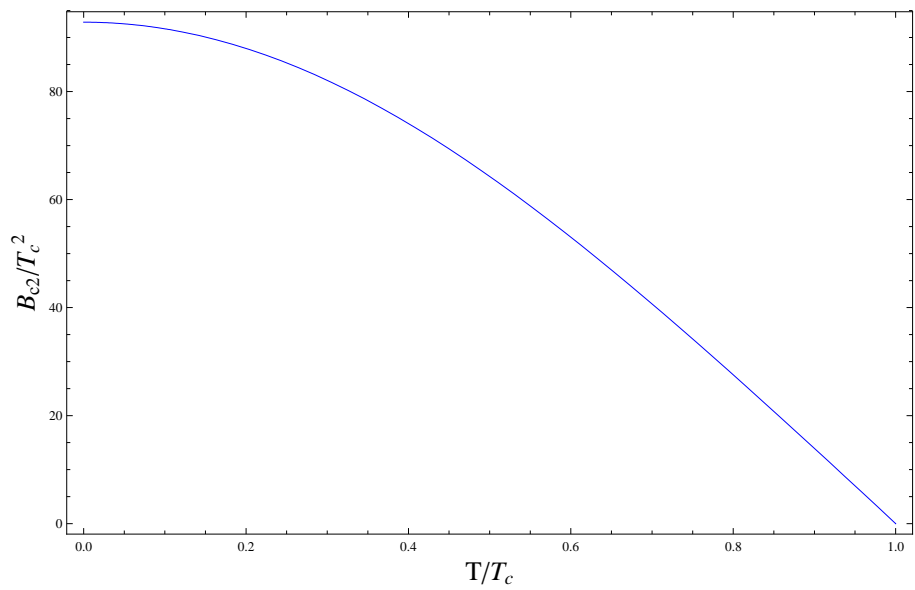}
\end{center}
\caption{ The upper critical magnetic field as a function of
$\frac{T}{T_c(0)}$. This figure indicates that stronger external
magnetic field $B$ leads to lower critical temperature $T$. $T_{0}$
denotes the critical temperature without external magnetic field.}
\label{one}
\end{figure}
By substituting $|\phi~'(1)|=\frac{\rho}{r_+}=\frac{3\rho}{4\pi
l^2T}$ into (\ref{BT}), we obtain the ansatz for $B_{c2}$ \be
B_{c2}=\frac{1}{9}\left(\sqrt{5376
\pi^4T^4+\frac{81\rho^2}{l^4}}-112\pi^2T^2\right) \ee The relation
between upper critical magnetic field $B_{c2}$ and the critical
temperature $T_c(0)$ can be determined by using (\ref{critical})
\be\label{Bc}
B_{c2}=\frac{16}{9}\pi^2T^2_c(0)\left(\sqrt{7}\sqrt{3\frac{T^4}{T^4_c(0)}+4}-7\frac{T^2}{T^2_c(0)}\right).
\ee From (\ref{Bc}), we can see that as a superconductor is cooled
down through the critical temperature $T_c(0)$, the critical field
gradually increases to its maximum value $B_{c2}$ at absolute zero
$T=0$. The temperature dependence of $B_{c2}(T)$ is exhibited in
Fig. {\ref{one}}. Similar figures can be found in \cite{ns,wen}.
Therefore, we can recover  the result in Ginzburg-Landau theory
where $B_{c2}\propto [1- T /{T_c(0)}]$ from the analytic approach.

\section{Holographic Superconductors immersed in external magnetic field with higher curvature corrections}

The spatially independent condensate solutions with higher curvature
corrections which corresponds to the superconducting phase below the
critical temperature was investigated in \cite{gre} and \cite{pw}.
With the presence of the external magnetic field, it is of great
interest to explore the spatially dependent condensate solutions and
further examine the influence of the higher curvature corrections on
the condensation.

We start with the neutral AdS black hole solution in $D$ dimensions
described by the metric \cite{g3,g4}
\begin{eqnarray}
\label{metric} d s^2 &=& \displaystyle -H(r)N^2d t^2+H^{-1}(r)d
r^2+\frac{r^2}{l^2} d x^{i}d x^{j},
\end{eqnarray}
with
\begin{eqnarray}
H(r)&=& \frac{r^2}{2
\tilde{\alpha}}\left[1-\sqrt{1-\frac{4\tilde{\alpha}}{l^2}\bigg(1-\frac{ml^2}{r^{D-1}}}\bigg)
\right],\nonumber\\&=&\frac{r^2}{2\lambda
l^2}\left[1-\sqrt{1-4\lambda\bigg(1-\frac{r^{D-1}_{+}}{r^{D-1}}
\bigg)}\right],\nonumber\\
 \Lambda &=&-\frac{(D-1)(D-2)}{2l^2},
\end{eqnarray}
where  $\tilde{\alpha}=(D-4)(D-3)\alpha'$, $\alpha'$ is a
Gauss-Bonnet coupling constant with dimension $\rm(length)^2$,
$\lambda=\tilde{\alpha}/l^2$ and the parameter $l$ corresponds to
the AdS radius. $N$ is a dimensionless constant which
specifies the speed of light of the boundary theory. The horizon is
located at $r=r_{+}$. The gravitational mass $M$ is expressed as
\begin{equation}
M=\frac{(D-2)V_{D-2}}{16 \pi G_D  }m.
\end{equation}
Taking the limit $\alpha'\rightarrow 0$, the solution reduces to the
Schwarzschild AdS black hole. The constant $N^2$ in the metric
(\ref{metric}) can be fixed at the boundary where the geometry
reduces to Minkowski metric conformally, i.e.\ $d s^2\propto -c^2d
t^2+d\vec{x}^2$. When $r\rightarrow\infty$, we have
$$
H(r)N^2 \rightarrow\frac{r^2}{l^2},
$$
so that $N^2$ is found to be
\begin{equation}
N^2=\frac{1}{2}\Big(1+\sqrt{1-4 \lambda}\ \Big).\label{N}
\end{equation}
Note that the boundary speed of light is specified to be unity
$c=1$.   The
black hole Hawking temperature is defined as
\begin{equation}\label{hawk}
T=\frac{1}{2\pi\sqrt{g_{rr}}}\frac{d \sqrt{g_{tt}}}{d
r}=\frac{(D-1)Nr_+}{4\pi l^2}.
\end{equation}
After fixing the horizon radius $r_{+}$ and the boundary speed
of light to be unity, the boundary theory temperature $T$ reaches
its minimum at $\lambda=\frac{1}{4}$ and goes to infinity as
$\lambda\rightarrow -\infty$.  In this background, we consider a
Maxwell field and a charged complex scalar field with the action
\begin{equation}
S=\int d^D x \sqrt{-g}\left[-\frac{1}{4}F^{\mu\nu}F_{\mu\nu}-|\nabla
\psi-iA\psi|^2-m^2 |\psi|^2\right].
\end{equation}
We assume that these fields are weakly coupled to gravity, so they
do not backreact on the metric satisfying the probe approximation.

\subsection{Weak magnetic field limit}
In the weak magnetic field limit, we can repeat the discussion on
the spatially independent condensate solutions reported in
\cite{gre,pw}.  The equations of motion of scalar and Maxwell fields
read
\begin{eqnarray}
&&z^{D-2}\partial_{z}\left[\frac{1}{z^{D-4}r^2_{+}}H(z)\partial_{z}\psi\right]+\frac{
A^2_t}{z^2H(z)N^2}\psi-\frac{m^2}{z^2}\psi=0,\label{main11}\\
&&
\partial^2_{z}A_t-\frac{D-4}{z}\partial_{z}A_t=\frac{2r^2_{+}}{z^4 H(z)}A_t|\psi|^2,
\end{eqnarray}
where $z=\frac{r_{+}}{r}$. Near the AdS boundary, the solutions
behave like \be \psi=c_{-}z^{\beta_{-}}+c_{+}z^{\beta_{+}},
~~~A_t\equiv\phi(z)=\mu-\frac{\rho}{r^{D-3}_{+}}z^{D-3} ,\ee where
$\beta_{\pm}=\frac{1}{2}\left[(D-1)\pm\sqrt{(D-1)^2+4m^2N^2l^2}\right]$.
We will set $c_{-}=0$ in the following and fix $l$ in the calculation. In the theory with the Gauss-Bonnet correction, the AdS curvature
length is given by $l_{AdS} = lN$, it seems more appropriate to fix $lN$ instead of $l$. However it was checked in [30] that fixing $l$ or $lN$ makes no difference in presenting the same qualitative features  as the Gauss-Bonnet factor varies.  Regularity at the horizon requires \be \psi(1)=-\frac{D-1}{m^2l^2}
\psi'(1),~~~\phi(1)=0.\ee
 Applying the analytic approach proposed
in\cite{gre,pw}, we can get the expectation value of the
2-dimensional operator $<\mathcal {O}>$\cite{pw}
\begin{eqnarray}
&&\frac{<\mathcal {O}>^{\frac{1}{\beta_{+}}}}{T_c}=\Upsilon
\frac{T}{T_c}\left[\left(\frac{T_c}{T}\right)^{D-2}
\left(1-\frac{T}{T_c}\right)^{D-2}\right]^{\frac{1}{2\beta_{+}}},\label{od}\\
&&\Upsilon=\frac{4\pi}{(D-1)N}\left\{{\sqrt{(D-1)[1+(4-D)(1-z_{m})]}
~[2(D-1)+m^{2}l^{2}(1-z_{m})]}\right\}^{\frac{1}{\beta_{+}}}\nonumber\\&&
\bigg\{\sqrt{2(1-z_{m})}(D-1)
[2z_{m}+(1-z_{m})\beta_+]z_{m}^{\beta_{+}-1}\bigg\}^{-\frac{1}{\beta_{+}}},
\end{eqnarray}
where the critical temperature is derived in the form
\begin{equation}\label{ot}
T_c(B=0)=\frac{(D-1)N}{4\pi l^2} \left(\frac{(D-3)z^{D-4}_m\rho
r_{+}}{\left[1+(4-D)(1-z_m)\right]|\phi'(1)|}\right)^{1/(D-2)}.
\end{equation}
It is clear that $<\mathcal {O}>$ is zero at $T=T_c$ and
condensation occurs for $T<T_c$. For positive Gauss-Bonnet coupling
constant, it is true that the critical temperature $T_c$ decreases
as $\lambda$ becomes more positive, which means that condensation
become harder to form for higher curvature corrections\cite{pw}.
Considering that the Gauss-Bonnet constant is also allowed to be
negative from causality, in Fig.\ref{11} we also plot the critical
temperature for negative Gauss-Bonnet coupling constant. The
condensates as functions of temperature by comparing the influences
due to positive, zero and negative Gauss-Bonnet coupling constants
are exhibited in Fig.\ref{2}. We see that different from the
positive Gauss-Bonnet coupling constant, the allowed negative
coupling can enhance the condensation and make the superconducting
phase easier to form.

 \begin{figure}[htbp]
 \begin{minipage}{1.1\hsize}
\begin{center}
\includegraphics*[scale=0.6] {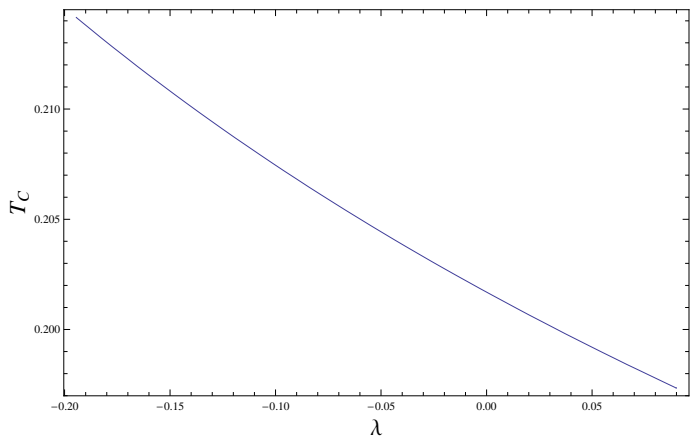}
\end{center}
\caption{ The critical temperature as a function of the Gauss-Bonnet
coupling constant. Here we set $D=5$, $\rho=1$, $m^2 l^2=-3$,
$r_{+}=1$ and $z_m=1/2$.} \label{11}
\end{minipage}
\end{figure}
 \begin{figure}[htbp]
 \begin{minipage}{1\hsize}
\begin{center}
\includegraphics*[scale=0.5] {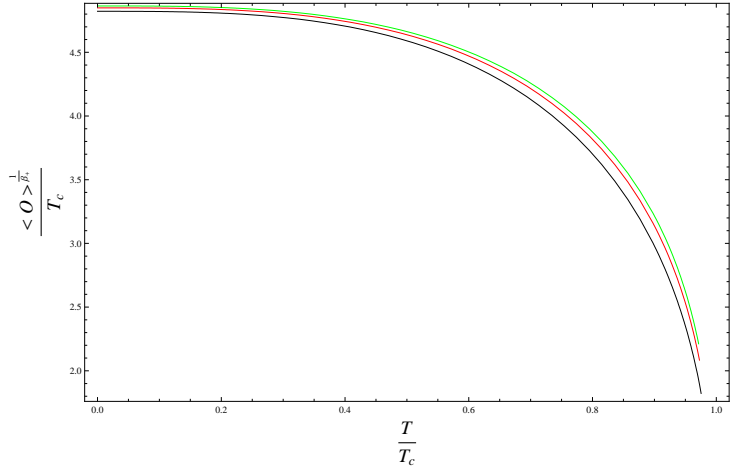}
\end{center}
\caption{(color online) The condensate as a function of temperature
for various values of $\lambda$. We fix the mass of the scalar field
$m^2l^2=-3$ in $D=5$ dimensions. The three lines from the bottom to
the top corresponds to $\lambda=-0.19$ (black), 0 (red) and $0.1$
(green), respectively.} \label{2}
\end{minipage}
\end{figure}

\subsection{Strong magnetic field limit}
When the magnetic field is strong enough, we have to consider the
spatially dependent solutions. It is consistent to take the ansatz
\be \psi=\psi(x,y,z),~~~A_t=\phi(z),~~~A_{i(i\neq
y)}=0,~~~A_y=B_{c2}x. \ee The main equation now governing the scalar
field $\psi$ is
\begin{equation}
z^{D-2}\partial_{z}\left[\frac{l^2}{z^{D-4}r^2_{+}}H(z)\partial_{z}\psi\right]+\frac{l^2
A^2_t}{z^2H(z)N^2}\psi-\frac{m^2l^2}{z^2}\psi=-\frac{1}{\alpha^2}\left(\partial^2_{i}\psi
+(\partial_y-iA_y)^2\right)\psi,
\end{equation}where $
\alpha=\frac{r_{+}}{l^2}$. We can separate
\begin{equation}
\psi=e^{ik_y y} X(x)R(z),
\end{equation}
where $X(x)$ satisfies the equation for a two dimensional harmonic
oscillator with frequency determined by $B_{c2}$ \be \label{hermite}
-\left(\partial^2_{x}
-(k_y-B_{c2}x)^2\right)X(x)={\lambda_{n}B_{c2}}X(x). \ee The
solution of (\ref{hermite}) is the Hermite function $H_n$ \be
X(x)=e^{-\frac{(B_{c2}x-k_y)^2}{2 B_{c2}}} H_n(x),\ee where
$\lambda_n=2n+1$ denotes the separation constant. We will choose the
lowest mode $n=0$ in the following of our computation. After the
separation of variables, we obtain the equation of motion for
$R(z)$,
\begin{equation}\label{Rz22}
R''+\frac{H'(z)}{H(z)}R'-\frac{(D-4)}{z}R'+\frac{r^2_{+}\phi^2}{N^2z^4H^2(z)}R
-\frac{m^2r^2_{+}}{z^4H(z)}R=\frac{B_{c2}l^2}{z^2H(z)}R(z).
\end{equation}
The regularity at the horizon gives
\begin{equation}
R'(1)=\left[-\frac{m^2l^2}{D-1}-\frac{B_{c2}}{(D-1)\alpha^2}\right]R(1).
\end{equation}
Near the AdS boundary, we have
\begin{equation}\label{asm}
R(z)=c_{+}z^{\beta_{+}}.
\end{equation}
 We will  use the notation
 $\phi(z)=\frac{\rho}{r^{D-3}_+} (1-z^{D-3})$
in the following computation. By expanding $R(z)$ in a Taylor series
near the horizon, we have
\begin{equation}\label{exp1}
R(z)=R(1)-R'(1)(1-z)+\frac{1}{2}R''(1)(1-z)^2+\ldots.
\end{equation}
Near $z=1$, (\ref{Rz22}) gives
\begin{eqnarray}
\label{R21}
R''(1)&=&-\left(m^2l^2+\frac{B_{c2}}{\alpha^2}\right)\lambda R(1)
+\frac{1}{2}\left(\frac{B_{c2}/\alpha^2+m^2l^2}{D-1}\right)^2R(1)\nonumber\\
&&+\frac{m^2l^2}{(D-1)}R(1)
-\frac{\phi^{'}(1)^2}{2N^2(D-1)^2\alpha^2}R(1),
\end{eqnarray}
where $\alpha=\frac{r_{+}}{l^2}$. The approximate solution of $R(z)$
near the horizon can be given by
\begin{eqnarray}\label{RR}
R(z)&=&R(1)+\left(\frac{m^2l^2}{D-1}+\frac{B_{c2}}{(D-1)\alpha^2}\right)R(1)(1-z)
\nonumber\\&+&\frac{1}{2}\bigg\{-\left(m^2l^2+\frac{B_{c2}}{\alpha^2}\right)\lambda
 +\frac{1}{2}\left(\frac{B_{c2}/\alpha^2+m^2l^2}{D-1}\right)^2\nonumber\\
&&+\frac{m^2l^2}{(D-1)} -\frac{\phi^{'}(1)^2}{2N^2(D-1)^2\alpha^2}
\bigg\}R(1)(1-z)^2.
\end{eqnarray}
Matching (\ref{asm}) and (\ref{RR}) at the intermediate point
$z_{m}$, we obtain
\begin{eqnarray}
c_+
z^{\beta_{+}}_{m}&=&R(1)+\left(\frac{m^2l^2}{D-1}+\frac{B_{c2}}{(D-1)\alpha^2}\right)R(1)(1-z_m)
\nonumber\\
&&+\frac{1}{2}\bigg\{-\left(m^2l^2+\frac{B_{c2}}{\alpha^2}\right)\lambda
 +\frac{1}{2}\left(\frac{B_{c2}/\alpha^2+m^2l^2}{D-1}\right)^2\nonumber\\
&&+\frac{m^2l^2}{(D-1)} -\frac{\phi^{'}(1)^2}{2N^2(D-1)^2\alpha^2}
\bigg\}R(1)(1-z_m)^2.\label{con1}\\
 \beta_{+}c_+
z^{\beta_{+}-1}_{m}&=&-\left(\frac{m^2l^2}{D-1}+\frac{B_{c2}}{(D-1)\alpha^2}\right)R(1)
-\bigg\{-\left(m^2l^2+\frac{B_{c2}}{\alpha^2}\right)\lambda
 +\frac{1}{2}\left(\frac{B_{c2}/\alpha^2+m^2l^2}{D-1}\right)^2\nonumber\\
&&+\frac{m^2l^2}{(D-1)} -\frac{\phi^{'}(1)^2}{2N^2(D-1)^2\alpha^2}
\bigg\}R(1)(1-z_m). \label{con2}
\end{eqnarray}
Using Eqs.(\ref{con1}) and (\ref{con2}), we can eliminate $c_{+}$
and get
\begin{eqnarray}
|\phi~'(1)|&&=\frac{N}{\alpha}\bigg\{B^2_{c2}(z_m-1)\left((\beta_{+}-2)z_m-\beta_{+}\right)
-2B_{c2}\alpha^2\bigg[((D-1)\lambda-m^2l^2)(\beta_{+}-2)z^2_{m}\nonumber\\
&&+2(\beta_{+}-1)z_{m}(D-1+m^2l^2-(D-1)^2\lambda)+\beta_{+}\left[2-m^2l^2
+(D^2+1)\lambda\right.\nonumber\\
&&\left.-2(1+\lambda)D\right]\bigg]
+\alpha^4\bigg[m^2l^2(z_m-1)\left((\beta_{+}-2)z_m-\beta_{+}\right)
+4(D-1)^2\beta_{+}\nonumber\\
&&-2(D-1)m^2l^2\bigg((D-1)\beta_{+}\lambda-3\beta_{+}
-2z_m(\beta_{+}-1)((D-1)\lambda-2)\nonumber\\
&&+z^2(\beta_{+}-2)((D-1)\lambda-1)\bigg)\bigg]\bigg\}^{1/2}
\bigg\{(z_m-1)\left((\beta_{+}-2)z_m-\beta_{+}\right)\bigg\}^{-1/2}.
\end{eqnarray}
 By using
$|\phi~'(1)|=\frac{(D-3)\rho}{r^{D-3}_{+}}$ and (\ref{hawk}), we
find the expression for the upper critical magnetic field $B_{c2}$,
which turns out to be very long and involving.  In order to have a
simple formula, let us first fix the parameters as $D=5$, $z_m=1/2$,
$l=1$, and $m^2=-3$.
\begin{eqnarray}
B_{c2}&=&\bigg\{2\bigg[2\pi^6T^4\bigg(\beta^2_{+}(3-64\lambda+32\lambda^2)
+4\beta_{+}(3-48\lambda+32\lambda^2)+4(11-32\lambda+32\lambda^2)\bigg)\nonumber\\
&+&(\beta_{+}+2)\rho^2/T^2\bigg]^{1/2}-\pi^3T^2(10-32\lambda+13\beta_{+}-16\beta_{+}\lambda)\bigg\}
/{\left(N^2\pi(\beta_{+}+2)\right)}
\end{eqnarray}
  By further using
equation (\ref{ot}), we finally obtain \bea
{B_{c2}}&=&\frac{\pi^2T^2}{N^2(2+\beta_{+})}
\bigg\{\bigg[8\bigg(\beta^2_{+}(3-64\lambda+32\lambda^2)
+4\beta_{+}(3-48\lambda+32\lambda^2)+4(11-32\lambda+32\lambda^2)\bigg)\nonumber\\
&+&(\beta_{+}+2)(192\lambda+96\beta_{+}\lambda+145\beta_{+}-126)T^6_c/T^6\bigg]^{1/2}
\nonumber\\
&-&(10-32\lambda+13\beta_{+}-16\beta_{+}\lambda)\bigg\}, \eea where
$\beta_{+}=2+\sqrt{4-3N}$.
\begin{figure}[htbp]
 \begin{minipage}{1\hsize}
\begin{center}
\includegraphics*[scale=0.6] {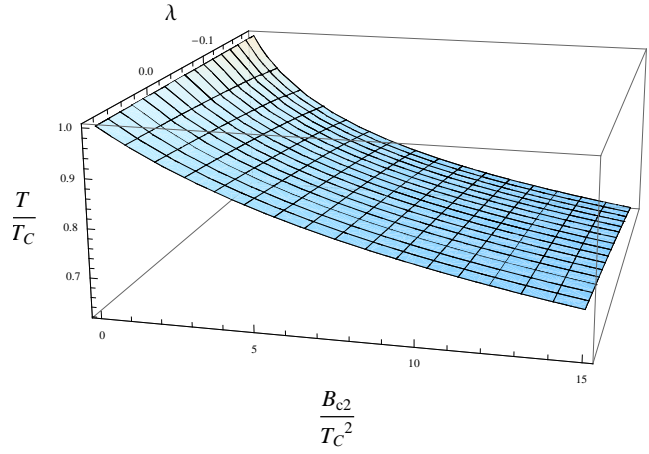}
\end{center}
\caption{(color online) The upper critical magnetic field against
$T/T_c$ and the Gauss-Bonnet coupling $\lambda$. $T_{c}$ denotes the
critical temperature without external magnetic field. We choose
$D=5$, $m^2l^2=-3$ and $z_m=1/2$ here.} \label{4}
\end{minipage}
\end{figure}
\begin{figure}[htbp]
 \begin{minipage}{1\hsize}
\begin{center}
\includegraphics*[scale=0.6] {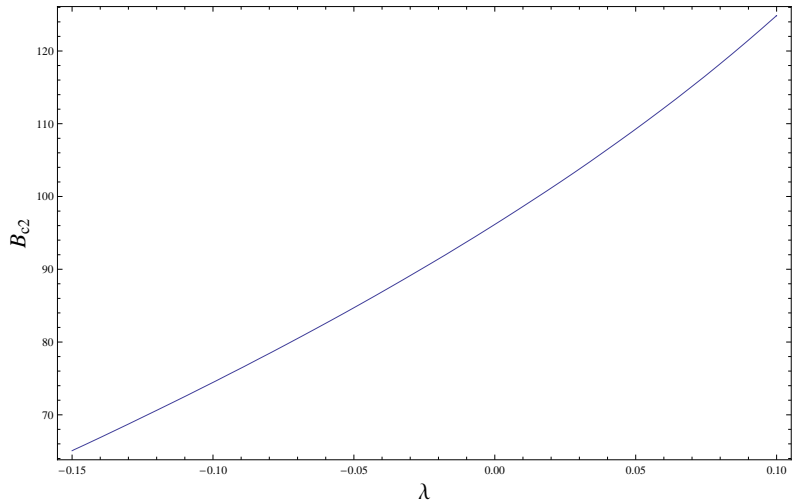}
\end{center}
\caption{(color online) The upper critical magnetic field as a
function of the Gauss-Bonnet coupling constant. We set $T=0.6$,
$T_{c}=1$, $D=5$, $m^2l^2=-3$ and $z_m=0.5$ here} \label{41}
\end{minipage}
\end{figure}
\begin{figure}[htbp]
 \begin{minipage}{1\hsize}
\begin{center}
\includegraphics*[scale=0.6] {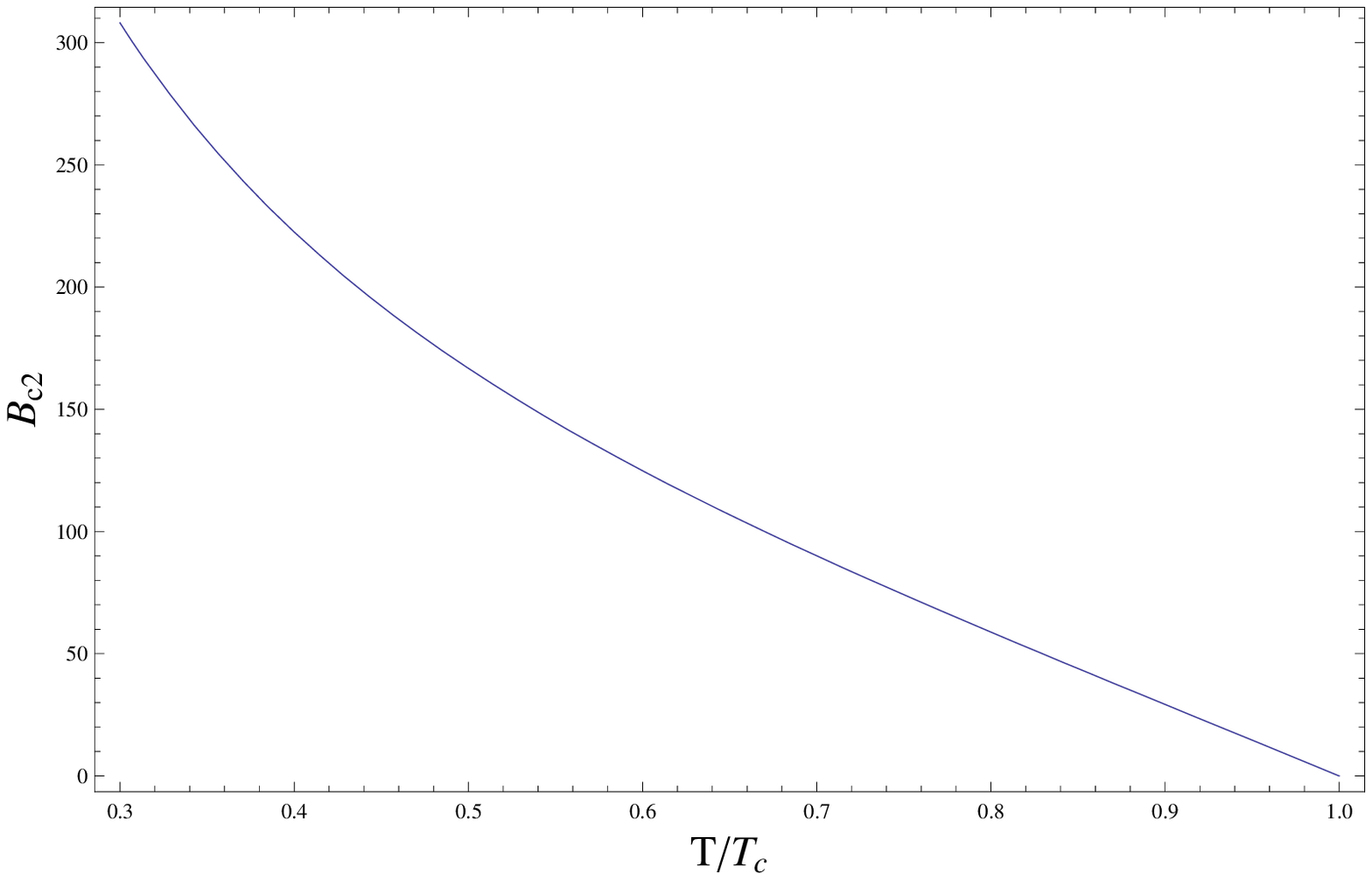}
\end{center}
\caption{(color online) The upper critical magnetic field as a
function of the temperature $T/T_{c}$ for fixed Gauss-Bonnet
coupling $\lambda=0.1$.  We set $m^2l^2=-3$, $D=5$, $T_c=1$ and
$z_m=0.5$ here.} \label{5}
\end{minipage}
\end{figure}
The expression for the upper critical magnetic field here is
different from that in four dimensional holographic superconductors. One
may find that as $T\rightarrow 0$, the critical magnetic field
$B_{c2}$ diverges, but it works well for $T\sim T_c$. This is true
for all $D\geq 5$ dimensional cases. There are several parameters in
the above equation, which relate to the critical magnetic field in
the presence of the Gauss-Bonnet coupling. Fixing the Gauss-Bonnet
coupling constant, we can see from Fig.\ref{4} that the critical
temperature $T$ decreases rapidly as the increase of the external
magnetic field. The decrease is even faster for negative
Gauss-Bonnet coupling, which shows that the magnetic field is expelled
strongly on the condensate for the negative Gauss-Bonnet coupling
situation. Fig.\ref{11}  and Fig.\ref{4} also exhibit that when the
magnetic field is not big enough, the influence given by the higher
curvature on the condensation keeps the same as that for neglecting
magnetic field: a more negative $\lambda$ leads to the higher
critical temperature $T$ while the bigger positive Gauss-Bonnet
coupling results in the smaller critical temperature.
 Fig.\ref{41} exhibits that for fixed temperature
$T<T_c$, a more negative $\lambda$ leads to the lower critical
magnetic field $B_{c2}$ while the bigger positive Gauss-Bonnet
coupling results in the bigger critical magnetic field. Fig.\ref{5}
shows us that higher temperature $T$ leads to lower critical
magnetic field $B_{c2}$.

\section{Conclusions and discussions}
In this work we have investigated the holographic superconductors
immersed in an external magnetic field by using the analytical
method developed in \cite{gre,pw}. We have obtained the spatially
dependent condensate solutions in the presence of the magnetism. We
have found analytically that the upper critical magnetic field
satisfies the relation $B_{c2}\propto (1-T/T_c)$, which is in
agreement with that given in the Ginzburg-Landau theory. Further we
found from the analytic approach that when the external magnetic
field grows, the condensation will be harder to form, which
indicates that the magnetic field expels the condensate as reported
 in \cite{aj,ns}.

We have further extended our investigation to the D-dimensional
Gauss-Bonnet AdS black holes and examined the influence given by the
Gauss-Bonnet coupling on the condensation. In the presence of the
magnetism we observed that the influence due to the positive
Gauss-Bonnet coupling keeps the same which will hinder the
condensation to form. Considering that the Gauss-Bonnet coupling
constant can also be negative without violating the causality on the
boundary theory, we have also examined the negative Gauss-Bonnet
coupling effect on the condensation. We found that different from
the positive coupling, the negative Gauss-Bonnet coupling can
enhance the condensation {when the external magnetic field is not
strong enough.} At the first glance the effect of the negative
Gauss-Bonnet coupling violates the Mermin-Wagner theorem. However to
examine whether or not the Mermin-Wagner theorem holds, one needs to
concentrate on the $4$-dimensional higher curvature gravity. Here we
have studied the Gauss-Bonnet gravity in dimensions $D\geq 5$, which
is non-trivial and ghost free. This may leave the space for the
enhancement of the condensation due to the negative Gauss-Bonnet
coupling. The similar relation between the upper critical magnetic
field with the temperature to that in the Ginzburg-Landau theory has
also been observed with the Gauss-Bonnet coupling.

\vspace*{10mm} \noindent
 {\large{\bf Acknowledgments}}

\vspace{1mm} We wish to thank  R. K. Su, Q. Y. Pan, X. He, S. Y.
Yin, Y. Peng and Y. Q. Liu for useful discussions. The work of BW
was supported by the NSFC. The work of XHG and SFW were partially
supported by NSFC under Grant Nos. 10947116 and 10905037, Shanghai
Education Development Foundation, and Innovation Foundation of
Shanghai University. XHG was also partly supported by Shanghai
Rising-Star program and SRF for ROCS, SEM. \vspace{1mm}


\renewcommand{\theequation}{A.\arabic{equation}}

\setcounter{equation}{0} \setcounter{footnote}{0}

\end{document}